\documentclass[acmsmall,nonacm]{acmart}
\makeatletter
\def\@titlefont{\fontsize{14.26}{16.83}\selectfont\sffamily\bfseries}
\makeatother

\usepackage{booktabs}   
\usepackage{listings}   
\usepackage{xcolor}

\definecolor{codekw}{RGB}{0,0,180}
\definecolor{codestr}{RGB}{163,21,21}
\definecolor{codecomment}{RGB}{0,128,0}

\lstdefinelanguage{TypeScript}{
  keywords={await, async, function, return, const, let, var, if, else, for, of,
    while, try, catch, finally, throw, new, class, interface, implements, extends,
    export, import, from, type, enum, public, private, readonly, void, this},
  keywordstyle=\color{codekw}\bfseries,
  morecomment=[l]{//},
  morecomment=[s]{/*}{*/},
  commentstyle=\color{codecomment}\itshape,
  morestring=[b]',
  morestring=[b]",
  morestring=[b]`,
  stringstyle=\color{codestr},
  sensitive=true,
}

\begin{document}

\title{Composable Building Blocks for Resilient Asynchronous Code}

\author{Frank Tip}
\affiliation{%
  \institution{Northeastern University and Amazon Web Services}
  \city{Boston}
  \state{Massachusetts}
  \country{USA}
}
\email{f.tip@northeastern.edu}

\begin{abstract}
Asynchronous calls to a network service, database, or language model must cope
with transient errors, slow or missing responses, throttling, and atomicity
violations. We show how higher-order combinators solve such problems uniformly,
including timeouts, retries, rate limiting, caching, reentrant locking, and
cancellation. Every combinator maps an async function to another of the same
type, so they share a uniform \emph{shape} and compose by nesting into one
expression that implements a program's whole resilience and concurrency
policy, leaving its business logic untouched. The same design spans both of
JavaScript's native async shapes, promise-returning and async-iterable-returning
functions, with one vocabulary of concerns. Solutions exist across the
ecosystem but are scattered over differently shaped libraries that are hard to
combine. We present case studies where the combinators are used to harden real
packages by adding missing resilience or concurrency control and replacing
bespoke policy.
\end{abstract}

\maketitle

\section{Introduction}
\label{sec:intro}

Consider an agent that decomposes a request into subtasks and calls a large
language model for each. The code is short and unremarkable:

\begin{lstlisting}
async function callLLM(prompt: string): Promise<string> {
  const res = await fetch("https://api.example.com/v1/complete", {
    method: "POST",
    body: JSON.stringify({ prompt }),
  });
  return (await res.json()).text;
}
async function runAgent(subtasks: string[]): Promise<string[]> {
  return Promise.all(subtasks.map((task) => callLLM(promptFor(task))));
}
\end{lstlisting}

\noindent
Against a responsive server and a few subtasks, this works, but production conditions expose several failure modes. Since \texttt{Promise.all} launches all requests at once, it can exceed the endpoint's rate limit and trigger \texttt{429 Too Many Requests} responses. Moreover, no individual call has a deadline, so a single slow or hung response stalls the whole batch. A lone transient \texttt{5xx} on a subtask also rejects the entire \texttt{Promise.all}, discarding
results of all siblings that already succeeded. Finally, if subtasks issue identical requests, the agent repeatedly pays for work it already completed.

Each of these failures has a standard, well-understood remedy: a timeout bounds a slow call, 
a retry absorbs a transient error, a rate limiter paces requests below the server's threshold, 
a concurrency limit caps the size of the burst, and a cache reuses previous results. 
While libraries exist that implement all of these remedies, they come in disparate shapes
such as wrapper functions, methods on policy objects, and classes that need to be 
instantiated. Assembling several around a single request means
reconciling their mismatched interfaces by hand, dissolving \texttt{runAgent} into
a tangle of timing, error-handling, and bookkeeping code that obscures the business logic.

The approach we propose took shape gradually as we
ran into these problems across our own projects~\cite{testpilot,llmorpheus}. It
removes the mismatch by modeling each remedy as a \emph{higher-order combinator}: a function that takes an async function and returns an async function of the same type, wrapping the
original with a single policy. Since combinators return the same type as their input, they compose naturally, and the hardened client is obtained by simply passing \texttt{callLLM} through a stack of them:

\begin{lstlisting}
const robustCallLLM = withCache(          // reuse results for identical calls
  withRetry(                              // up to 3 attempts
    withRateLimit(                        // >= 200 ms between calls
      withTimeout(callLLM, 10_000),       // 10 sec deadline for each call
      200),
    3, { delayMs: 500, backoff: "exponential" }));
\end{lstlisting}

\noindent
Only one line of the agent changes: \texttt{callLLM} becomes
\texttt{robustCallLLM}. This cleanly separates \emph{what} the agent does, which
stays in \texttt{runAgent} and \texttt{callLLM}, from \emph{how} it withstands an unreliable network, now handled entirely by the combinators. Their uniform, composable shape is what enables this
separation, and it is the central idea of this article.

Concurrency and cancellation raise additional concerns. JavaScript and
TypeScript provide no built-in concurrency control (unlike, say, Java's
\texttt{synchronized} blocks), yet concurrent asynchronous functions can 
interleave and exhibit race conditions and atomicity violations, which can be
prevented using combinators. When an agent revises its plan or a user
interrupts an operation, work already in flight must be stopped. Cancellation
is subtler than it looks: a cancelled call has not failed, so it
must propagate through a stack of combinators rather than being
retried or masked as an error. Handling these concerns consistently becomes
considerably more complex for stream-based APIs, since a lock may need to be
held for a stream's entire lifetime, and cancelling
a task mid-stream must be reconciled with whatever partial results the stream
has already produced.

Why does this matter now, and in this ecosystem? These problems have long
existed wherever a program depends on an external system, but LLM-based
agents bring them to the foreground at a new scale: a single agent may issue
hundreds of model calls, each slow, rate-limited, and prone to transient
failure. Solutions vary across ecosystems:
.NET and Java converged on comprehensive resilience libraries,
Polly~\cite{polly} and resilience4j~\cite{resilience4j}, whereas JavaScript
and TypeScript have no comparable convergence. A comprehensive library,
cockatiel~\cite{cockatiel}, exists, but it neither supports mutual exclusion
nor extends to stream-based APIs, and most of the ecosystem instead relies on
many small, single-purpose packages.

This article makes four contributions:
\begin{itemize}
  \item \textbf{A uniform design.} Every asynchronous concern is expressed as a
    higher-order combinator that maps an async function to an async function
    of the same type, so that policies compose by nesting and the business logic
    stays free of them.
  \item \textbf{Support for both of JavaScript's native async shapes.} Each concern
    has a promise-returning and an async-iterable-returning combinator, sharing
    the same name and composing the same way, so one vocabulary of concerns
    covers both shapes.
  \item \textbf{A treatment of cancellation.} We separate inbound cancellation, a
    caller cancelling a pending wait, from outbound cancellation, a computation
    signalling that it was cancelled, and unify them so that cancellation
    propagates cleanly through a stack of combinators.
  \item \textbf{A reentrant lock combinator.} We define a combinator built
    on \texttt{AsyncLocalStorage} to track per-chain ownership, letting a
    computation re-enter a lock it holds without deadlocking, which existing
    JavaScript locks cannot do without the deprecated \texttt{domain} module.
\end{itemize}
The combinators are available as an open-source package,
\texttt{async-combinators}~\cite{async-combinators}; Section~\ref{sec:case-studies}
applies them to existing packages as case studies.

\section{A Catalogue of Combinators}
\label{sec:catalogue}

Table~\ref{tab:catalogue} lists the full set of combinators, grouped by the kind
of problem each addresses. Each combinator has a promise-returning and an
async-iterable-returning variant, sharing the same name. Implementing the
streaming variants raises subtleties of its own, discussed in
Section~\ref{sec:streaming}. The first three groups act on a single call; the
last coordinates calls that run concurrently. This section describes each
combinator. Section~\ref{sec:challenges} discusses challenges that arise 
in their implementation.

\begin{table}[t]
  \caption{A catalogue of asynchronous combinators, grouped by theme.}
  \label{tab:catalogue}
  \vspace*{-3mm}
  {\footnotesize
  \begin{tabular}{lll}
    \toprule
    \textbf{Theme} & \textbf{Problem} & \textbf{Combinator} \\
    \midrule
    Failures       & transient failures            & \texttt{withRetry} \\
                   & service unavailable           & \texttt{withFallback} \\
    \midrule
    Scheduling     & requests taking too long      & \texttt{withTimeout} \\
                   & throttling                    & \texttt{withRateLimit} \\
    \midrule
    Nondeterminism & memoization                   & \texttt{withCache} \\
                   & record/replay                 & \texttt{withRecordReplay} \\
    \midrule
    Concurrency    & atomicity violations          & \texttt{withLock}, \texttt{withReentrantLock} \\
                   & too many concurrent requests  & \texttt{withMaxConcurrency} \\
    \bottomrule
  \end{tabular}
  }
\end{table}

\subsection{Failures: \texttt{withRetry} and \texttt{withFallback}}

Calls that cross a process boundary may experience \emph{transient} failures. 
A network request, database query, or file operation may error because of a momentary condition. In such cases,  simply reissuing the call may resolve the problem.
\texttt{withRetry} retries a failed call up to a fixed number of attempts, spacing them by
an optional exponential backoff (with jitter) so that a struggling server is not
overwhelmed and many clients do not retry in lockstep. When a failure is not
transient, because the service is down or the attempts are exhausted,
\texttt{withFallback} instead replaces the error with an alternative: a default
value, a previously recorded result, or a call to a secondary service.

\subsection{Scheduling: \texttt{withTimeout} and \texttt{withRateLimit}}

A call that never returns is as much a problem as one that fails. A server may
have crashed or  respond unacceptably slowly, and a caller is usually
better served by a prompt failure than an unbounded wait. \texttt{withTimeout}
rejects a call that does not settle within a deadline. The opposite concern is
calling a service \emph{too} often: many APIs cap the request rate  to
protect their infrastructure or meter usage, and a client that exceeds the cap
is met with a \texttt{429 Too Many Requests} response.
\texttt{withRateLimit} spaces successive calls by a fixed minimum interval.

\subsection{Nondeterminism: \texttt{withCache} and \texttt{withRecordReplay}}

Many asynchronous APIs are nondeterministic, yet programs often need repeated
calls with the same arguments to return the same result, to avoid redundant
work and stay consistent. \texttt{withCache} memoizes results within
a single run, storing the pending promise rather than the resolved
value to collapse concurrent duplicate calls into one in-flight request.
\texttt{withRecordReplay} extends this across runs by persisting results to disk. 
This enables reliable integration tests that run offline and are
free of a live service's nondeterminism.

\begin{lstlisting}
async function getExchangeRate(from: string, to: string): Promise<number> {
  const res = await fetch(`https://api.example.com/rate?from=${from}&to=${to}`);
  return (await res.json()).rate;
}
// Set RECORD=1 to refresh fixtures against the real service.
const rate = withRecordReplay(getExchangeRate, "fixtures/rates", {
  mode: process.env.RECORD ? "record" : "replay",
});
\end{lstlisting}

\subsection{Concurrency: \texttt{withLock}, \texttt{withReentrantLock}, \texttt{withMaxConcurrency}}

JavaScript does not support parallelism at the language level, but \texttt{async} functions can be interleaved at \texttt{await} expressions. Hence, computations that read
shared state and later write it back can experience atomicity violations.
Concurrent updates to an account balance are the classic case:

\begin{lstlisting}
async function updateBalance(id: string, delta: number): Promise<void> {
  const balance = await store.get(id);   // read
  await store.set(id, balance + delta);  // ...then write
}
await Promise.all([updateBalance("a", +10), updateBalance("a", -5)]);
\end{lstlisting}

\noindent
Here, both calls execute \texttt{store.get} and read the starting balance, 100, before either
calls \texttt{store.set}. The deposit then writes 110 and the withdrawal writes 95, so
that the deposit is lost and the final account balance is 95. \texttt{withLock} restores atomicity by protecting \texttt{updateBalance} with a single shared lock, so that one call's read and write
cannot interleave with the other's.

\noindent
\texttt{withLock} suffices when a guarded operation such as \texttt{updateBalance} 
stands alone, but real operations often rely on smaller ones that also need
to be guarded. A composed operation attempting to re-acquire a non-reentrant
lock it already holds will wait for that very lock, deadlocking against
itself. \texttt{withReentrantLock} allows a call chain to re-enter a lock it
already holds, so composed operations proceed while independent chains stay
serialized. Section~\ref{sec:cs-locking} presents a case study illustrating this.
Finally, \texttt{withMaxConcurrency} addresses a
related but distinct need, bounding how many calls run at once rather than enforcing
mutual exclusion.

\section{Implementation Challenges}
\label{sec:challenges}

We begin by looking at one combinator in detail to convey the idea, then turn
to the concerns that make a few of them subtle to implement.
Stripped of input validation, backoff timing,
and cancellation, all of which we return to below, \texttt{withRetry} looks as follows:

\begin{lstlisting}
function withRetry<ArgTypes extends any[], RtrnType>(
  fn: (...args: ArgTypes) => Promise<RtrnType>,
  maxAttempts: number
): (...args: ArgTypes) => Promise<RtrnType> {
  return async (...args: ArgTypes): Promise<RtrnType> => {
    for (let attempt = 1; ; attempt++) {
      try { return await fn(...args); } 
      catch (err) { if (attempt >= maxAttempts) throw err; }
    }
  };
}
\end{lstlisting}

\noindent
The two type parameters capture the wrapped function's shape:
\texttt{ArgTypes} is its argument tuple, and \texttt{RtrnType} the type its
promise resolves to, so \texttt{fn} is an arbitrary async function returning exactly that shape. The body of \texttt{withRetry} is
otherwise unremarkable: it returns a new function that forwards its arguments to \texttt{fn} and
re-invokes it until the call succeeds or attempts run out.

\subsection{Composition and Ordering}
\label{sec:composition}

Each combinator returns a function of the same type, so the output of one is a valid input to the next, and TypeScript infers the
argument and result types through the whole stack.
Composing combinators is achieved by nesting function calls, and the nesting
order is significant: each combinator wraps the next, so the order decides
how the policies interact.
Placing \texttt{withTimeout} inside \texttt{withRetry} gives every attempt its own
deadline, whereas placing it outside imposes one deadline on the entire retry loop:

\begin{lstlisting}
withRetry(withTimeout(fetch, 1_000), 3);   // 3 attempts, each capped at 1s
withTimeout(withRetry(fetch, 3), 1_000);   // 3 attempts, 1s for all of them
\end{lstlisting}

\noindent
The same reasoning generalizes to longer chains of nested combinators: retries
nested inside a rate limiter are themselves paced by it, and a cache placed
outermost returns a hit without invoking any of the policies beneath it. As we will
discuss next, composition also imposes an obligation on each
combinator to forward a caller's signal to the function it wraps, so that an abort reaches the
operation running at the bottom of the stack.

\subsection{Cancellation}
\label{sec:cancellation}

Cancellation has two aspects that are easy to conflate. \emph{Inbound} cancellation
is a caller abandoning a call that is still running or waiting to start. By
convention the caller passes an \texttt{AbortSignal} in a trailing
\texttt{\{ signal \}} argument, as \texttt{fetch} does. A combinator that makes the
caller wait for a lock, concurrency slot, or retry backoff watches this
signal and, when it aborts, drops the queued or pending work and rejects, rather
than running it later or leaking it:

\begin{lstlisting}
const controller = new AbortController();
const fetchLimited = withMaxConcurrency(fetchProfile, 4);
const p = fetchLimited(id, { signal: controller.signal });
controller.abort();   // if this call is still queued, it is dropped, never run
\end{lstlisting}

\noindent
\emph{Outbound} cancellation happens when a function reports itself as cancelled,
by throwing an \texttt{AbortError} (what an
\texttt{AbortController} produces). The combinators must not mistake
this for a failure: a cancelled call is not retried by \texttt{withRetry},
masked by \texttt{withFallback}, or stored by \texttt{withCache} or
\texttt{withRecordReplay}. The combinators detect it by name (\texttt{error.name} is
\texttt{"AbortError"}), a test that holds across execution contexts, and the
two aspects unify through that same name, since an inbound abort surfaces as an
\texttt{AbortError} on the way out. \texttt{withTimeout} rejects with a distinct \texttt{TimeoutError}, which the resilience combinators treat as an ordinary, retryable failure.

\subsection{Reentrant Locking}
\label{sec:reentrancy}

\texttt{withReentrantLock} poses an implementation
problem the other combinators do not. To let a call chain re-enter a lock it
already holds, the lock must be able to tell whether the code now asking to acquire
it \emph{is} that chain. A threaded language answers this by recording the owning
thread (e.g., via Java's \texttt{Thread.currentThread()}), but JavaScript has no
threads and no thread identity. We recover an
equivalent notion from \texttt{AsyncLocalStorage}, which associates state with an
asynchronous call chain and preserves it across the chain's \texttt{await}s. The
lock records a token identifying its current holder there, so acquiring it
first checks whether the current chain already holds that same token:

\begin{lstlisting}
async runExclusive<T>(fn: () => Promise<T>): Promise<T> {
  if (this.holder.getStore() === this.ownerToken) return fn();  // already holds it: re-enter
  const token = await this.acquire();         // otherwise wait for the lock, get a fresh token
  try { return await this.holder.run(token, fn); } // mark this chain the holder while fn runs
  finally { this.release(token); }
}
\end{lstlisting}

\noindent
The reentrant fast path, when the stored token matches the lock's owner,
runs \texttt{fn} without queuing. An unrelated chain finds no match, waits
for the lock, then runs \texttt{fn} inside \texttt{holder.run(token, $\cdots$)}
so its own nested acquisitions re-enter in turn. Reentrancy is uncommon among JavaScript lock libraries and the
alternatives fall short (Section~\ref{sec:related}). \texttt{AsyncLocalStorage}
is the supported, modern mechanism for chain-scoped state, which makes it the
natural basis.

\subsection{Streaming Semantics}
\label{sec:streaming}

\begin{sloppypar}
Extending the catalogue to \texttt{AsyncIterable}-returning functions raises
subtleties promise-returning combinators never face, because a stream can
fail \emph{mid-stream}, after having delivered items.
\texttt{withRetry} illustrates this well: retrying a failed promise costs
nothing, since the caller has seen no result, but retrying a stream that
already yielded items would replay consumed output or, for a
nondeterministic source (e.g., an LLM's token stream), diverge entirely. A
failure before the first item retries transparently, but a mid-stream failure is
retryable only if the caller asserts that \texttt{fn} is deterministic (with
the \texttt{resumable: true} option). Stripped of backoff and cancellation, the loop is:
\end{sloppypar}

\begin{lstlisting}
let delivered = 0;
for (let attempt = 1; attempt <= maxAttempts; attempt++) {
  let seen = 0;
  try {
    for await (const item of fn(...args)) {
      if (seen++ < delivered) continue;  // resuming: skip the already-seen prefix
      delivered++;
      yield item;
    }
    return;                                    // stream finished normally
  } catch (err) {
    if (attempt >= maxAttempts) throw err;
    if (delivered > 0 && !resumable) throw err; // partial output, not resumable: give up
  }
}
\end{lstlisting}

\noindent
The wrapper trusts \texttt{resumable} rather than verifying it, since
comparing items would require buffering the stream. It catches one
violation for free: a resumed run yielding \emph{fewer} items than
delivered, which throws instead of silently truncating.
A second subtlety is caching: since a stream has no single settled value to
share, \texttt{withCache} instead buffers items and lets each consumer replay
from its own cursor, sharing one upstream pull among all consumers waiting
for it.

\section{Case Studies}
\label{sec:case-studies}

We present three case studies in which we apply the combinators to real, widely-used software.

\subsection{Making HTTP requests resilient}
\label{sec:cs-fetch}

The first case study is focused on \texttt{fetch}, a popular API for making HTTP
requests. Native \texttt{fetch} offers no retry, no request timeout, no rate
limiting, and no caching. Resilience is left entirely to the caller. Fetching from an API is therefore exposed to transient failures, rate limits, and slow responses.
\texttt{fetch} resolves its promise even for error statuses: a \texttt{429} or
\texttt{503} arrives as a fulfilled \texttt{Response}, not a rejection. For
\texttt{withRetry} to treat those as retryable, the call must throw on them:

\begin{lstlisting}
async function checkedFetch(url: string): Promise<Response> {
  const res = await fetch(url);
  if (res.status === 429 || res.status >= 500) throw new Error(`HTTP ${res.status}`);
  return res;
}
const getItem = (id: number) => checkedFetch(`${base}/items/${id}`).then((r) => r.json());
\end{lstlisting}

\noindent
With that in place, and because the combinators share one shape, the full policy
for a service that is both flaky and rate-limited is a single composed
expression:

\begin{lstlisting}
const robustGetItem = withRetry(                  // retry transient failures
  withRateLimit(                                  // pace requests under the rate limit
    withTimeout(getItem, 5_000),                  // impose deadline on each attempt
    200),
  3, { delayMs: 500, backoff: "exponential" });
\end{lstlisting}

\noindent
The application code is untouched: \texttt{getItem} still just calls \texttt{fetch},
and the resilience lives entirely in the combinators. The same wrapped fetch can be
handed to a client such as \texttt{openapi-fetch} through its
\texttt{fetch} option, hardening every request it makes without changing the
client. Moreover, \texttt{withCache} and \texttt{withMaxConcurrency}
compose in the same way when deduplication or a bound on the number of concurrent in-flight requests is wanted. A few combinators thus add the robustness a program needs, without
having to resort to a heavier HTTP client for its built-in resilience. 

\subsection{Coordinating concurrent access to a JSON database}
\label{sec:cs-locking}

Our second case study is focused on \texttt{lowdb}, a widely used local JSON database
(22{,}000+ GitHub stars). It loads data into memory and, on every
\texttt{write()}, serializes it back to a file. It
lacks concurrency control, and its documentation points users to a
database such as PostgreSQL if concurrency control is required. The obvious way
to update a balance, illustrated by the \texttt{deposit} function below, involves
reading it, doing some asynchronous work, then writing it. Run
concurrently, all deposits read the same
starting balance before any of them writes, and updates are lost:

\begin{lstlisting}
async function deposit(db, id: string, amount: number): Promise<void> {
  const balance = db.data.accounts[id] ?? 0;    // read (missing account starts at 0)
  await work();                                 // e.g. validation or a network call
  db.data.accounts[id] = balance + amount;      // modify
  await db.write();                             // persist the whole database
}
await Promise.all([10, 20].map((amt) => deposit(db, "alice", amt))); // update lost, alice = 20
\end{lstlisting}

\noindent
This is the atomicity problem of Section~\ref{sec:catalogue} manifesting in a
real package. The locking combinators address it without modifying
\texttt{lowdb} at all: we introduce a factory function that builds a set of
operations sharing a \emph{reentrant} lock, wrapping each with
\texttt{withReentrantLock} so that a composed operation can re-enter the lock
through the lower-level operations it calls:

\begin{lstlisting}
function makeBank(db: Low<Data>) {
  const lock = new ReentrantLock();
  const getBalance = withReentrantLock(async (id: string) => {
    return db.data.accounts[id] ?? 0; }, lock);
  const deposit = withReentrantLock(async (id: string, amount: number) => {
    await setBalance(id, (await getBalance(id)) + amount); }, lock);
  // setBalance, withdraw similar
  const transfer = withReentrantLock(async (from: string, to: string, amount: number) => {
    await withdraw(from, amount);
    await deposit(to, amount); }, lock);
  return { getBalance, setBalance, deposit, withdraw, transfer };
}
\end{lstlisting}

\noindent
Concurrent deposits on an account are now serialized, so no updates are lost:

\begin{lstlisting}
const bank = makeBank(db);
await Promise.all([10, 20].map((amt) => bank.deposit("alice", amt)));
await bank.transfer("alice", "bob", 25); // alice = 5, bob = 25; atomic and no deadlock
\end{lstlisting}

\noindent
The composition works because the lock is reentrant: \texttt{deposit} and
\texttt{transfer} re-enter it through the lower-level operations they call, whereas a
non-reentrant lock would deadlock. This gives users a practical middle ground: where
lowdb's own guidance is to migrate to a full database once concurrency becomes a
concern, a reentrant lock supplies exactly the missing atomicity, letting a
project keep the simplicity of a local JSON file until it genuinely outgrows it.

\subsection{Making an LLM-based agent resilient}
\label{sec:cs-strands}

\begin{sloppypar}
The third case study applies stream combinators to Strands
(\texttt{strands-agents/harness-sdk}), a widely used TypeScript SDK for LLM
agents (6{,}800+ GitHub stars). Its only resilience
is a custom retry mechanism that retries \texttt{ModelThrottledError} with
exponential backoff for up to six attempts. There is no timeout and no rate
limit at the model layer.
Strands' streaming interface is a natural fit for the stream combinators. The
entry point is its \texttt{Model.stream} method:
\end{sloppypar}

\begin{lstlisting}
abstract stream(messages: Message[], options?: StreamOptions): AsyncIterable<ModelStreamEvent>
\end{lstlisting}

\noindent
This matches the streaming combinator shape exactly, and wrapping
\texttt{OpenAIModel.stream} with \texttt{withTimeout} and \texttt{withRateLimit}
was straightforward: requests are now paced, and any that stall past the
deadline fail retryably.

\begin{sloppypar}
Replacing Strands' own retry mechanism with \texttt{withRetry} was more involved,
due to how its agent loop is built. Each cycle dispatches lifecycle events
(\texttt{BeforeModelCallEvent}, \texttt{AfterModelCallEvent}) on an event bus
that tracing, metrics, and user hooks observe, and its retry is \emph{inversion
of control}: a hook sets \texttt{event.retry = true} on
\texttt{AfterModelCallEvent} to trigger another attempt. By contrast, the combinators assume
\emph{direct control}: \texttt{withRetry(fn, n, opts)} owns its
own loop and calls \texttt{fn} up to \texttt{n} times internally. Wrapping
\texttt{OpenAIModel.stream} in the full stack,
\texttt{withRetry(withTimeout(withRateLimit(stream)))}, works functionally, but
since the loop emits exactly one \texttt{AfterModelCallEvent} per cycle
regardless of how many attempts happened inside \texttt{stream()}, retry
counts and per-attempt timing never reach the event bus: three retried
attempts would surface as one abnormally slow call.
\end{sloppypar}

To fix this, we refactored Strands' hardcoded retry loop into an \texttt{attempt} function
that dispatches Before/After events for one try and throws on failure,
plus a pluggable driver that wraps \texttt{attempt} and decides whether to
retry. The default driver reproduces the historical loop exactly but an
application can instead pass \texttt{withRetry} itself as the driver, through a new
\texttt{modelCallDriver} option, while disabling Strands' own mechanism. Because
\texttt{withRetry} now wraps the emitting \texttt{attempt} rather than
\texttt{stream()} itself, every retry, error, and timing reaches the event bus.
A \texttt{shouldRetry} predicate reproduces Strands' original retry
decisions, extended to also treat \texttt{TimeoutError} as retryable, so the
combinator-based loop handles both throttling and timeout failures:

\begin{lstlisting}
const driver = config.modelCallDriver ?? defaultDriver   // default = today's loop
return yield* driver(attemptModelCall)
\end{lstlisting}

\noindent
An application can now supply a combinator as the driver:

\begin{lstlisting}
new Agent({ model,
            retryStrategy: null,
            modelCallDriver: (attempt) => withRetry(attempt, 3, { shouldRetry }), })
\end{lstlisting}

\noindent
Overall, this required only a small, targeted change to Strands (the
\texttt{modelCallDriver} seam and \texttt{shouldRetry} predicate) in exchange
for resilience built from reusable, independently testable primitives, without
sacrificing Strands' native per-attempt observability.

Runnable versions of all three case studies are linked from the
\texttt{async-combinators} repository~\cite{async-combinators}; the fetch and
lowdb studies are self-verifying via assertions, while the Strands study is an
interactive demo whose combinator activity is logged for direct observation.
\section{Related Work}
\label{sec:related}

Resilience is a well-studied topic. In the .NET ecosystem, Polly~\cite{polly} has
offered mature, comprehensive mechanisms for it, packaging retry, timeout,
circuit breaking, rate limiting, fallback, and hedging behind a common interface.
Polly decorates the \emph{execution} of an operation rather than the operation
itself: a resilience pipeline is a separate object through which a delegate is
run, so composing several strategies means building and executing through that
pipeline object. Its rate limiter, delegating to .NET's
\texttt{System.Threading.RateLimiting}, supports sliding-window and token-bucket
algorithms that reject or queue calls exceeding a permit limit. By contrast, our
\texttt{withRateLimit} combinator paces every call to a fixed minimum interval and
never rejects one outright.

In the Java ecosystem, resilience4j~\cite{resilience4j} takes an approach closer
to ours. Decorators such as \texttt{Retry.decorateSupplier} and the chained
\texttt{Decorators.ofSupplier(...)\allowbreak.withRetry(...)\allowbreak.withCircuitBreaker(...)\allowbreak.decorate()}
builder wrap a \texttt{Supplier} and return a new one, much like our combinators
wrap a function and return a new one. Each resilience4j decorator depends on
Java's \texttt{Supplier}/\texttt{CompletionStage} types and its own
configuration object (\texttt{RetryConfig}, \texttt{CircuitBreakerConfig}, ...)
rather than one uniform shape, and its \texttt{Bulkhead}, like Polly's
concurrency limiter, is a resource-isolation mechanism, not the mutual-exclusion
primitive our reentrant lock provides. Its circuit breaker offers capabilities we
do not: sliding windows, half-open recovery, and per-strategy event streams for
observability. For streams, \texttt{resilience4j-reactor} extends the same
decorators to Project Reactor's~\cite{reactor} \texttt{Flux}, a reactive type
built on Reactive Streams.

Our contribution, relative to Polly and resilience4j, is not the individual
mechanisms for adding resilience, but the manner of composing them into one
uniform, signature-preserving shape (Section~\ref{sec:catalogue}), with no
policy object, builder, or per-strategy configuration type to learn, and,
unlike either library, a reentrant lock for enforcing mutual exclusion between
asynchronous operations.

Within the JavaScript ecosystem, \texttt{cockatiel}~\cite{cockatiel} offers a
comprehensive catalogue modeled closely on Polly: retry, circuit breaking,
timeout, bulkhead, and fallback, all invoked through a policy object's
\texttt{execute(callback)} method. The same
concerns are otherwise addressed by many single-purpose packages. The popular
\texttt{p-*}
family~\cite{p-retry,p-timeout,p-limit,p-memoize} supplies retries, deadlines,
concurrency limiting, and memoization as separate modules. Moreover,
\texttt{async-mutex}~\cite{async-mutex} and \texttt{async-lock}~\cite{async-lock}
provide locking. These libraries do not share a shape: \texttt{p-retry} and
\texttt{p-memoize} wrap a function, \texttt{p-timeout} wraps an
already-created promise, \texttt{p-limit} returns a limiter function used to
wrap other functions, and \texttt{async-mutex} and \texttt{async-lock} are
classes exposing methods such as \texttt{runExclusive} or \texttt{acquire}
that must be invoked explicitly. \texttt{opossum}~\cite{opossum} similarly
bundles circuit breaking, timeout, and fallback in a stateful class invoked
via \texttt{fire()}, sharing no shape with our catalogue. Assembling several of
these mechanisms means reconciling mismatched interfaces by hand, the
fragmentation our work sets out to remove. Reentrant locking is a further gap: \texttt{async-mutex}
is non-reentrant, and \texttt{async-lock} offers reentrancy only through Node's
deprecated \texttt{domain} module, which motivates the \texttt{AsyncLocalStorage}-based
lock of Section~\ref{sec:reentrancy}.

For streams specifically, JavaScript has its own pair of reactive-extensions
libraries. RxJS~\cite{rxjs} applies operators such as \texttt{retry},
\texttt{timeout}, and \texttt{throttleTime} to its own \texttt{Observable} type,
a push-based abstraction distinct from the language's native
\texttt{AsyncIterable}. IxJS~\cite{ixjs} wraps native \texttt{AsyncIterable}
and \texttt{Iterable} values directly, but still requires adopting its own
wrapper class and chained-operator style rather than plain functions.
Neither library's \texttt{retry} distinguishes deterministic from
nondeterministic sources: both simply resubscribe to (RxJS) or re-iterate
(IxJS) the source from the beginning on error, re-emitting whatever it produced
before the failure. Our stream \texttt{withRetry} instead defaults to letting
a failure that occurs after output has already been delivered propagate
unchanged, since silently replaying a nondeterministic source, a language
model's token stream chief among them, would splice two different
continuations together. Its \texttt{resumable: true} option opts
into the restart-and-skip behavior only for sources asserted to be
deterministic.

Various solutions record a program's external interactions and replay them
for deterministic, offline tests, an idea originating with VCR~\cite{vcr}'s
``cassette'' model. In JavaScript, \texttt{nock}~\cite{nock} records and
replays HTTP traffic, and for language-model calls tools such as
aimock~\cite{aimock} record and replay fixtures. Each works by intercepting
one specific transport layer: nock hooks into Node's HTTP module, and aimock
intercepts calls to an LLM API. \texttt{withRecordReplay} is
deliberately more general and lower-level: it records and replays the result
of \emph{any} async function in process, keyed on its arguments, without
intercepting a transport at all. That generality matters most for sources
with no dedicated transport layer, such as a database driver or a computed
result with no network call at all. Because \texttt{withRecordReplay} shares
the same shape as the other combinators, it composes with them directly:
wrapping it around an already-hardened
\texttt{withRetry(withTimeout(withRateLimit(fn)))} means recording begins
only once the call succeeds, so a fixture reflects the clean result, not the
transient failures a live service might encounter.

\section{Discussion and Conclusion}
\label{sec:discussion}

Expressing every asynchronous concern, resilience, concurrency, and
cancellation alike, as a single, signature-preserving higher-order combinator
enables the creation of a unified resilience policy by nesting one combinator
call inside another, while its business logic remains untouched. The same
approach applies to both of JavaScript's native async shapes, promise-returning
and async-iterable-returning functions, with one vocabulary of concerns. The
three case studies (Section~\ref{sec:case-studies}) show this working against
real, unmodified packages.

The uniform shape keeps the catalogue open for extension. Since each concern is handled using the same kind of function-to-function combinator, additional ones should compose with the existing set without additional complexity. Circuit breaking and
request hedging are the clearest candidates: Polly already implements both,
and resilience4j implements circuit breaking (Section~\ref{sec:related}), so
adding them here would not be a novel mechanism, but would bring them into the
same uniform vocabulary as the rest of the catalogue. Debouncing is another
concern that would fit the same mold. RxJS and IxJS (Section~\ref{sec:related})
already offer it as an operator over their reactive types, though not as a
plain function-to-function combinator.

\begin{sloppypar}
The combinators and all three case studies are available as open-source from \url{https://github.com/neu-se/async-combinators}. The
\texttt{async-combinators}~\cite{async-combinators} package is available on npm, and each case study is available in a separate repository, linked from the package's README.
\end{sloppypar}

\section*{Acknowledgments}

This work was supported in part by the National Science Foundation under Grant CCF-2307742.
Frank Tip is a Professor of Computer Science at Northeastern University and an Amazon Scholar.
The paper describes work performed at Northeastern University and is not associated with Amazon.

\bibliographystyle{ACM-Reference-Format}
\bibliography{references}

\end{document}